\documentclass[aps,prx,twocolumn,letterpaper,superscriptaddress]{revtex4-2}
\usepackage{graphicx}
\usepackage{amsmath}
\usepackage{amssymb}
\usepackage{url}
\usepackage{hyperref}

\begin{document}
\title{Wegner model in high dimension: $\mathrm{U}(1)$ symmetry breaking and a\\ non-standard phase of disordered electronic matter,\\
I. One-replica theory}
\author{Martin R. Zirnbauer}
\affiliation{Institut f\"ur Theoretische Physik, Universit\"at zu K\"oln, Z\"ulpicher Str. 77a, 50937 K\"oln, Germany}
\date{\today}

\begin{abstract}
The Anderson transition between localized and metallic states is traditionally analyzed by assuming a one-parameter scaling hypothesis. Although that hypothesis has been confirmed near two dimensions by $\varepsilon = d-2$ expansion of the Wegner-Efetov nonlinear $\sigma$-model, there exists mounting evidence that the transition in $d \geq 3$ may have a second branch and that two relevant parameters are needed in order to describe the universal behavior at criticality. Doubt of the standard hypothesis also comes from field theory. Indeed, increasing the space dimension pushes the Anderson transition towards strong disorder, where a strong-coupling approach very different from the usual weak-coupling analysis of the $\sigma$-model is called for. In the present work, we develop a novel field theory of Anderson transitions at strong coupling based on the key observation that the $\mathrm{U}(1)$ symmetry which distinguishes retarded from advanced fields may undergo spontaneous symmetry breaking. That symmetry breakdown splits the $\sigma$-model coupling into two, thus leading to a natural scenario of two-parameter scaling. While we develop the field theory from the concrete starting point of the Wegner $N$-orbital model, we believe our results to be of much wider applicability. The first of a series, the present paper offers a pedagogical introduction to the main ideas in the setting of the one-replica theory. Subsequent papers will employ the self-consistent approximation of Abou-Chacra \emph{et al.} and develop the full supersymmetric theory. The latter establishes the existence of a new renormalization-group fixed point, whose basin of attraction constitutes a third phase of disordered electronic matter.
\end{abstract}

\maketitle

\section{Introduction}

Mobile electrons in a disordered semiconductor exhibit a quantum critical phenomenon referred to as the Anderson transition \cite{PWA-1958} -- a quantum phase transition from a metallic state at weak disorder to an insulating state at strong disorder. In the present article, we revisit the field-theoretical description of the Anderson transition for non-interacting (or mean-field) electrons.

Following an influential paper \cite{GangOfFour}, it became standard to assume for the Anderson transition a one-parameter scaling hypothesis. The latter was substantiated by the renormalization group treatment in $d = 2 + \varepsilon$ dimensions of an effective field theory, the nonlinear $\sigma$-model of Wegner \cite{Weg79a} and Efetov \cite{Efe83}, which is derivable in controlled fashion for the metallic regime at weak disorder. For small $\varepsilon = d-2$, the Anderson-transition critical point lies at a weak field-theory coupling, where it can be studied by perturbation expansion; see \cite{EversMirlin} for a review.

Later it was argued \cite{MF94} that the nonlinear $\sigma$-model captures the critical behavior of Anderson transitions even in high dimension, the upper critical dimension being $d = \infty$. If so, one expects one-parameter scaling, with the dimensionless conductivity or local conductance as the only relevant parameter, and one predicts two (and no more than two!) phases: (i) the metallic state with absolutely continuous energy spectrum and spatially extended eigenfunctions, and (ii) the insulating state with pure point spectrum and localized eigenfunctions. The informed skeptic \cite{Goldenfeld}, however, will caution that other possibilities have not been ruled out. In fact, taking the field-theory perspective, the Anderson critical point moves into the strong disorder (hence strong coupling) regime with increasing dimension and there exists no proof that the Wegner-Efetov nonlinear $\sigma$-model (while perturbatively renormalizable) continues to be renormalizable in the non-perturbative regime at strong coupling. In the present article, we shall argue that the standard $\sigma$-model scenario for Anderson transitions in high dimension $d$ (meaning $d \geq 3$) needs revision and that a third phase, demonstrably distinct from the two known phases, may intervene in a suitable region of parameter space.

After this blunt overture, let us slow down and offer some background, history, motivation, and perspective for our work. In recent years, the venerable subject of Anderson transitions experienced a revival of interest, motivated mostly by questions arising from the topical area of many-body localization: in an effort to  shed light on the challenging and controversial situation with many-body localization, researchers revisited the old subject of Anderson transitions on tree-like graphs with high connectivity. Assuming such a setting, it was proposed \cite{KAI2018} that in addition to the two known phases (metallic and insulating) there may exist a third phase, called NEE (for non-ergodic extended), where the electron eigenstates are neither metallic nor insulating, but have (multi-)fractal characteristics \cite{AGKL1997}. Defying traditional theory, this proposal was borne out for various simplified models of random-matrix type \cite{KKCA2015, NKK19, RussDoll, NEE-Poisson} but has apparently been refuted for the Anderson model on tree-like graphs \cite{TMS2016, BHT2022, Lemarie2022, SLS2022}.

One clear (if somewhat surprising) outcome of these investigations is that the Anderson transition on high-dimensional random graphs does not conform to the traditional hypothesis of one-parameter scaling; rather, the large-scale numerical simulations done by several groups \cite{Lemarie2022, SLS2022} indicate that \underline{two} relevant parameters come into play. What remains controversial are the exact details of the critical behavior and the specific two-parameter scaling theory to adopt. Ref.\ \cite{Lemarie2022} focuses on the insulator side of the transition and argues for behavior of Kosterlitz-Thouless (KT) type. In contrast, Ref.\ \cite{SLS2022} focuses on the metal side and finds two length scales with power-law criticality (not KT). We take this unresolved controversy as a strong indication that an analytical approach is called for. In that vein, the present paper will go beyond the standard nonlinear $\sigma$-model, by adapting the Wegner-Efetov treatment of the metallic regime to high-dimensional (or high-mobility) systems with strong disorder. The final outcome of our analysis will be a field theory with two independent couplings, which naturally explains the observation of two-parameter scaling.

Truth be told, the original motivation for the present work came actually from a different source. A few years ago, the author proposed a conformal field theory \cite{CFT-IQHT} for the scaling limit of the critical point between plateaus of the integer quantum Hall (IQH) effect. That proposal broke with traditional thinking. While most of the finite-size scaling analysis that exists for the IQH transition is based on the postulate of a nonlinear $\sigma$-model in its variant due to Pruisken and collaborators \cite{Pruisken83}, the work of \cite{CFT-IQHT} starts from the observation that the nonlinear $\sigma$-model (NL$\sigma$M), as a description of the renormalization-group (RG) fixed point at criticality, conflicts with the principle of conformal invariance for RG-fixed points in two dimensions. (In short, 2D conformal invariance dictates \cite{Affleck85, CFT-IQHT} that if there is a conserved current, $\partial_\mu j^\mu = 0$, then the dual current must also be conserved: $\partial_\mu \epsilon^{\mu}_{\; \nu} j^\nu = 0$. A conflict arises because NL$\sigma$M gives the first conservation law but fails to deliver the second one.) The issue was resolved by abandoning NL$\sigma$M and constructing for the IQH critical point a theory compatible with conformal invariance, with a novel scenario of partial symmetry breaking as its key ingredient. That scenario is robust, i.e., not restricted to the IQH transition or even two dimensions. Hence, we may expect it to occur for a variety of Anderson transitions at strong coupling (meaning strong disorder and/or a small local density of states). As we shall see in the present paper, the partial symmetry breaking scenario is naturally associated with the appearance of a second critical length. Thus it transcends the traditional field-theory paradigm in the same way as the proposed NEE-phase transcends the traditional Anderson metal-insulator paradigm.

What is the physical interpretation of our scenario of partial symmetry breaking (PSB)? To answer that, we note that the real-space wave function is related to the field-theory mapping by a kind of Fourier duality: (i) when the wave function is delocalized, filling the position space more or less uniformly, the field-theory map is totally localized, clustering around a single point (selected by spontaneous symmetry breaking) of the target space. Conversely, (ii) when the wave function is localized in position space, the field-theory map is totally delocalized, fluctuating very strongly and roaming all over the target space (so that the symmetry of the field theory is fully restored). Now, what happens in our partial symmetry breaking scenario is this: the field-theory map is delocalized along the directions of a parallelizable submanifold of the target space, but is localized about the zero cross section of the corresponding normal bundle; simply put: the field spreads along some ``light-like'' directions of the target space but gets stuck in the transverse ``space-like'' directions. (The former effect entails symmetry restoration for a subgroup of the symmetry group, while the latter amounts to symmetry breaking for all group elements acting transversally.) By the stated principle of Fourier duality, our novel field-theory scenario translates to the following physical picture for a phase distinct from the two known phases of metal and insulator: (iii) the real-space wave function does not fill all of the position space, not even after any amount of coarse graining (that's the flip side of subgroup symmetry restoration); at the same time, the wave function fails to be localized (that's the flip side of transverse symmetry breaking) but explores a vanishing fraction of the infinite position space, most likely in the form of a percolating cluster.

In view of the above, we are intrigued by a very recent preprint \cite{Delande23} where precisely such a physical picture is advocated for the standard Anderson model in three dimensions. Employing the semiclassical tool of ``localization landscape'' (which is a poor man's version of real-space renormalization in that it replaces the rugged bare random potential by a smoother effective random potential) the authors of \cite{Delande23} argue that the 3D Anderson transition features a second branch (apparently unnoticed in previous work on the 3D Anderson model). Located near the periphery of the disorder-energy phase diagram, that branch separates localized states from wave functions that percolate but still lie outside (!) the known phase boundary \cite{Schreiber95} to the metallic states. This striking observation begs the question: how can the traditional picture with only two phases accommodate those percolating states that are neither localized nor metallic? (Our answer is that it cannot.) We take that conundrum as evidence that our strong-coupling two-parameter field theory for Anderson transitions in high dimension may apply in dimension as low as $d = 3$.

For completeness, let us add that there exists still another motivation for our work. From spectral theory one knows that a random Schr\"odinger operator (or, for that matter, any self-adjoint operator) in the infinite-volume limit has a spectral decomposition into three parts, which are called pure point, singular continuous, and absolutely continuous \cite{ReedSimon}. Roughly speaking, these spectral types correspond to localized, fractal, and fully extended eigenfunctions, respectively. The occurrence of pure point and absolutely continuous spectrum is well understood, but what about singular continuous spectrum \cite{WhatisLoc, DJLS96}? While singular continuous spectrum is naturally associated with critical states, one can ask whether there might exist a bigger stage for it. (The reader may be intrigued to learn that singular continuous spectrum has been proved to be generic in a certain sense \cite{DJMS94}.) Is it conceivable that partial symmetry breaking, the proposed NEE-phase, and singular continuous spectrum (more precisely, an RG-stable H\"older class thereof), could simply be different manifestations of the same thing? See \cite{AK2023} for some hints.

To summarize the main message of this introduction, there is now mounting evidence that the traditional one-parameter scaling picture for Anderson transitions is incomplete. Indeed, it has been discovered that the energy eigenstates in a certain region of parameter space are neither localized nor delocalized in the usual sense of a metal, but rather delocalized in a ``non-ergodic'' way, by forming a percolating cluster that covers only a vanishing fraction of the position space. To the extent that the latter feature is stable under renormalization by coarse graining, such wave functions constitute a separate phase (distinct from a region of crossover \cite{TMS2016, BHT2022} into the metallic phase). The main goal of the present paper is to firmly establish the novel third phase (``NEE'' or ``PSB'' or ``SC'') as a true thermodynamic phase in the sense of Landau, by developing a field-theory scenario in which a global symmetry is spontaneously broken (if only partially so) by the formation of an order parameter.

To avoid logical gaps and uncontrolled approximations, we shall derive the field theory starting from a variant of the standard Anderson model, namely the Wegner $N$-orbital model \cite{Weg79b} without time-reversal invariance, a.k.a.\ of symmetry type $A$ in the Tenfold Way \cite{suprev, HHZ}. Specifically, specializing to the band center for simplicity, we are going to show that the third phase materializes for sure when $N \gg 1$ and $w_V \gg \sqrt{N} w_T$ with $w_V$ the on-site disorder strength and $w_T$ the disordered hopping strength. (The first inequality indicates mobile electrons with semiclassical dynamics, the second one strong fluctuations of the on-site random potential.) Notwithstanding its specific starting point, we expect the outcome of our field-theoretical analysis to describe the critical behavior of a wide class of models, thanks to the principle of universality of critical phenomena.

Let us conclude this extended introduction with  some words on technique. The Wegner model has a probability law invariant under local $\mathrm{U}(N)$ gauge transformations, a property not shared by the more popular Anderson model (at the defining lattice scale) but expected to emerge under renormalization \cite{PS82}. The local $\mathrm{U}(N)$-invariance invites the introduction of matrix variables by a bosonization step \cite{BEKYZ07}, replacing the standard technique of Hubbard-Stratonovich transformation, which is inappropriate in the strong-coupling situation of interest. (Recall that our focus is on high space dimension, where the Anderson transition takes place in a regime of strong disorder.) The target space of the resulting matrix-field theory (more precisely, its boson-boson sector) is foliated by the hyperbolic orbits of a noncompact symmetry group. It should be stressed that one has to deal with a \emph{family} of symmetry orbits, not with a single such orbit as assumed by the usual nonlinear $\sigma$-model approximation.

Our first key observation is that the dominant contributions to the matrix-field integral come from a null-orbit on which the invariant metric (similar to the Minkowski metric for an on-shell massless relativistic particle) degenerates. In a bit more detail, the target-space geometry is highly anisotropic so that the field degrees of freedom split into strongly fluctuating variables tangential to a light ray on the null-orbit, and weakly fluctuating variables for the transverse directions. The metric anisotropy entails the coexistence of two widely different length scales (leading to the phenomenology of two-parameter scaling), and it opens the striking possibility for one coupling of the field theory to renormalize to zero while the other coupling flows to infinity, thus setting the foundation for our scenario of partial symmetry breaking. Building on the present paper's heuristic reasoning within the one-replica theory, a separate paper \cite{AZ23} looks for confirmation of the conjectured features by solving the one-orbital Wegner model in the approximation given by the self-consistent theory of localization \cite{AAT1973}.

Alas, it turns out that the $N = 1$ Wegner model does \underline{not} support a third phase! Hence, in a third publication \cite{AZ24} we push further to analyze the model for $N > 1$ on a $d$-dimensional lattice, by combining super-bosonization, gradient expansion, and approximate real-space renormalization based on the exact heat kernel of the reduced symmetry group. Following this, our second key insight is that the symmetry restoration for the unbroken subgroup (characteristic of the third phase) is a fully non-perturbative process driven by topological excitations in the fermion-fermion sector (present only for $N \ge 2$) of the supersymmetric field theory. Thus, the appearance of the PSB phase requires the cooperation of two agents: strong diagonal disorder imposes the near-degenerate Minkowski geometry of the boson-boson sector, while semiclassical electron mobility switches on the subgroup-symmetry restoring force of the fermion-fermion sector (traditionally believed to be irrelevant at the Anderson transition in high dimension \cite{MF94}). In the band center of the Wegner model, both ingredients are verifiably active in the parameter range of $1 \ll N \ll w_V^2 / w_T^2$ (and for a wider range away from the band center).

The outline is as follows. First of all, in Sect.\ \ref{sect:2}, we present our main message as a conjectured RG flow diagram for Anderson transitions at strong coupling. Then, in Sect.\ \ref{sect:3}, we present the Wegner model and set up its one-replica treatment by Fourier-Laplace transform (Sect.\ \ref{sect:3.A}). Discarding the commonly used Hubbard-Stratonovich transformation, we introduce a matrix-valued field $Q$ by a form of bosonization, initially so for $N = 1$ (\ref{sect:3.B}). Next, we focus our attention on the strong-disorder limit (Sect.\ \ref{sect:4}), where the dominant orbit of the symmetry group is a light cone in Minkowski space. To circumvent the light-cone singularity, we abandon the usual parametrization of the matrix $Q$ based on diagonalization. After adjusting the parametrization of $Q$ (Sect.\ \ref{sect:4.A}), we succeed in integrating out the weakly fluctuating non-Goldstone degrees of freedom (Sect.\ \ref{sect:4.B}). The resulting field theory is discussed for its anisotropy of couplings and compared with the standard $\sigma$-model in Sect.\ \ref{sect:4.C}. The generalization to the case with $N > 1$ orbitals is sketched in Sect.\ \ref{sect:4.D}. We then revisit (in Sect.\ \ref{sect:5}) the RG flow diagram and point out some robust predictions. Sect.\ \ref{sect:6} contains our conclusions.

\section{Conjectured RG Flow Diagram}\label{sect:2}

\begin{figure}
    \centering
    \includegraphics[width=8cm]{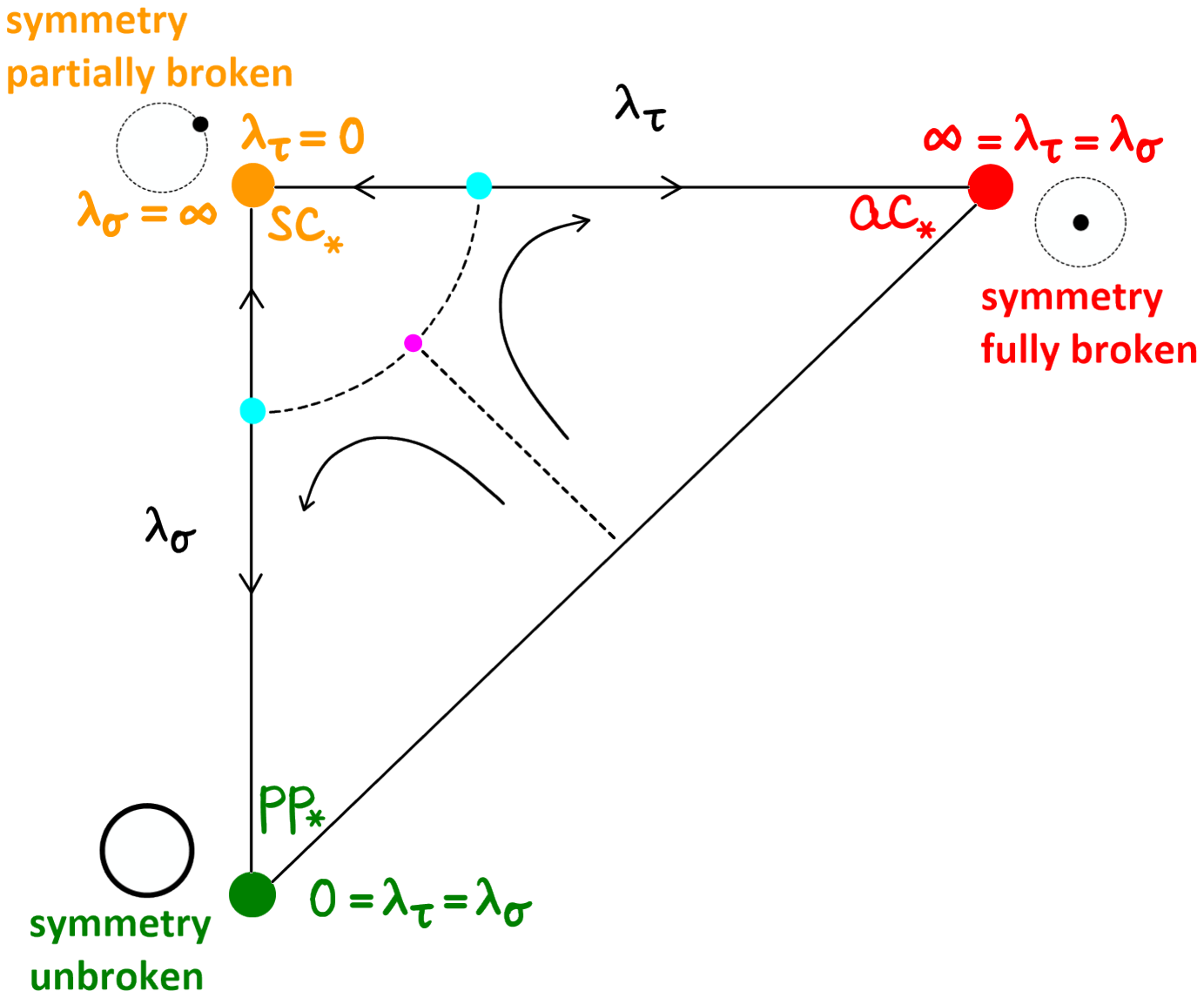}
    \caption{Schematic renormalization group (RG) flow diagram (conjectured) for the two-parameter field theory governing Anderson transitions at strong coupling ($d \geq 3$). The reciprocal coupling $\lambda_\tau$ (resp.\ $\lambda_\sigma$) is the stiffness for field fluctuations tangent to a ``time-like'' subspace of the target space (resp.\ transversal to that subspace, or ``space-like''). In addition to the two known RG-fixed points ``${\rm ac}_\ast$'' ($\lambda_\sigma = \lambda_\tau = \infty:$ metal) and ``${\rm pp}_\ast$" ($\lambda_\sigma = \lambda_\tau = 0:$ insulator), there is a third attractive RG-fixed point, tentatively labeled as ``${\rm sc}_\ast$'', at $\lambda_\sigma = \infty$, $\lambda_\tau = 0$. In a certain parameter range, the Wegner model for more than one orbital per site ($N > 1$) is argued to renormalize to ${\rm sc}_\ast$. For each of the three attractive RG-fixed points, the figure gives an indication of the statistically relevant field configurations projected to the Poincar\'e disk of the boson-boson sector: at ${\rm ac}_\ast$ the field clusters near the center of the disk, at ${\rm pp}_\ast$ it uniformly populates the boundary of the disk, and at ${\rm sc}_\ast$ it clusters near a point on the disk boundary.} \label{fig:F1}
\end{figure}

We summarize the main outcome of the present work in the form of a conjectured renormalization-group (RG) flow diagram (Fig.\ \ref{fig:F1}) for the field theory to be derived. The field-theory action functional in continuum approximation is presented in Eq.\ (\ref{eq:S-cont}) below. We emphasize that it differs from the standard Wegner-Efetov nonlinear $\sigma$-model but is still invariant under the infinitesimal action of $\mathrm{U}(1,1|2)$, known to be the symmetry group for disordered electrons of symmetry class $A$; cf.\ \cite{EversMirlin} .

The field theory features two couplings ($\lambda_\sigma \,, \lambda_\tau$), both with the meaning of a field stiffness: $\lambda_\tau$ for the field fluctuations tangential to a ``light-like'' subspace of the target space, and $\lambda_\sigma$ for the transverse or ``space-like'' fluctuations. The bare values of these couplings for the Wegner model are given in Eq.\ (\ref{eq:coupfin}). By setting $\lambda_\sigma = \lambda_\tau$ one recovers the standard $\sigma$-model. However, for the regime of strong disorder we find the inequality $\lambda_\sigma \gg \lambda_\tau$.

The undisputed backbone of the flow diagram are two trivial RG-fixed points: ${\rm pp}_\ast$ (for pure point spectrum of the random Hamiltonian) where $\lambda_\sigma = \lambda_\tau = 0$; and ${\rm ac}_\ast$ (for absolutely continuous spectrum) where $\lambda_\sigma = \lambda_\tau = \infty$. These fixed points are attractive for systems (at a given energy $E$) with localized and metallic states, respectively. The key feature of the diagram is a third attractive RG-fixed point, tentatively denoted as ${\rm sc}_\ast$, which resides at $\lambda_\tau = 0$ and $\lambda_\sigma = \infty$. This fixed point comes about because the boson-boson sector of the field theory causes symmetry breaking (due to $\lambda_\sigma \to \infty$) in the direction of the space-like field fluctuations, while the fermion-fermion sector (effectively present for $N \geq 2$) drives symmetry restoration (due to $\lambda_\tau \to 0$) by field diffusion along the time-like target space directions. Demonstrating the existence of ${\rm sc}_\ast$ is our main research goal.

Given the attractive fixed points ${\rm pp}_\ast$ and ${\rm ac}_\ast$, and taking ${\rm sc}_\ast$ for granted, the remainder of the flow diagram is more or less determined by basic principles, especially the principle of analyticity of the vector field generating the RG flow (the RG beta ``function'').  There must be three phase boundaries: pp/ac, ac/sc, and sc/pp. The most natural scenario is that these meet at a multi-critical point located in the interior of the triangle of $0 \leq \lambda_\tau \leq \lambda_\sigma < \infty$. Our second key input is that the line $\lambda_\sigma = \lambda_\tau$ is unstable under renormalization, in the strong-disorder situation of Anderson transitions in high dimension. If so, the RG flow near the multi-critical point is ``in to'' for the pp/ac phase boundary and ``out of'' for the ac/sc and sc/pp phase boundaries. Also by naturality, one expects the latter to terminate at the boundary of the triangle of stiffness parameters. Dictated by analyticity of the generating vector field, the  gross features of the RG flow are then as indicated in Fig. \ref{fig:F1}.

We note that when crossing the pp/ac phase boundary near the unstable line $\lambda_\sigma = \lambda_\tau$, one observes just one critical point (of course). Nonetheless, the near-critical RG flow does feel the presence of the intervening sc-phase. Indeed, the presence of ${\rm sc}_\ast$ and its basin of attraction separates two unstable RG-fixed points, one at $\lambda_\sigma = \infty$ and the other at $\lambda_\tau = 0$. The large-scale critical behavior on the ac-side of the pp/ac transition is governed by the former, whereas the behavior on the pp-side is governed by the latter. Such a difference in critical behavior has apparently been seen in numerical simulations of the Anderson model on a variety of random graphs \cite{Lemarie2022, SLS2022}. Moreover, a very recent study of the Anderson model on a three-dimensional cubic lattice \cite{Delande23} argues by semiclassical means that the pp/ac phase boundary bifurcates into two branches. We are inclined to identify these two branches with our ac/sc and sc/pp phase boundaries.

For completeness, we mention that a phenomenological renormalization group for the Anderson model on random regular graphs (modeling the situation for $d \to \infty$) has been proposed in \cite{VAKS23}. Let us therefore stress that the present paper addresses the case of a finite dimension $d \geq 3$. The discussion of the proposed RG flow diagram (Fig.\ \ref{fig:F1}) will be resumed in Sect.\ \ref{sect:5} at the end of the paper.

\section{Model and Technique Used}\label{sect:3}

The object studied in this paper is a random Hamiltonian, $H$, of Schr\"odinger type in the discrete setting of an infinite graph, $\mathbb{G}$, e.g.\ a cubic lattice $\mathbb{G} = \mathbb{Z}^d$. Each vertex (or site) of $\mathbb{G}$ hosts $N$ states, which are called ``orbitals''. The Hamiltonian is written as a sum
\begin{equation}\label{eq:Wegner}
    H = \sum_{n,n^\prime; \alpha\alpha^\prime} h_{n\alpha; n^\prime \alpha^\prime} \, c_{n \alpha}^\dagger c_{n^\prime \alpha^\prime}^{\vphantom{\dagger}}
\end{equation}
over orbital indices $\alpha = 1, \ldots, N$ and the sites $n$ of $\mathbb{G}$. Although $H$ is presented here in basis-independent form using particle creation ($c^\dagger$) and annihilation operators ($c$), we will analyze it as an operator ($h$) on the single-particle Hilbert space $\ell^2(\mathbb{G}) \otimes \mathbb{C}^N$. All Hamiltonian matrix elements are taken
\footnote{This choice is made only for simplicity, not by necessity; we could also handle a box or similar distribution, as we are going to use the flexible tool of bosonization instead of decoupling by a Hubbard-Stratonovich Gaussian field.}
to be uncorrelated complex Gaussian random variables (subject to the Hermiticity condition $h_{n\alpha; l \beta} = \overline{h_{l \beta ; n \alpha}}$) with zero mean and variance
\begin{align}\label{eq:problaW}
     \mathbb{E} \left( h_{n\alpha; n^\prime \alpha^\prime} h_{l \beta ; l^\prime \beta^\prime} \right) &= \delta_{n l^\prime} \delta_{n^\prime l} \delta_{\alpha \beta^\prime} \delta_{\alpha^\prime \beta} \cr &\times \left( \delta_{n n^\prime} w_V^2 + A_{n n^\prime} w_T^2 \right)
\end{align}
where $A_{n n^\prime} = A_{n^\prime n} \in \{ 0 , 1 \}$ is the adjacency matrix of the graph $\mathbb{G}$. The parameter $w_V$ sets the scale of variation of the on-site random matrix elements ($n=n^\prime$), while $w_T$ sets the kinetic-energy scale for hopping between adjacent sites ($n \not= n^\prime$). Officially known as the Wegner $N$-orbital model \cite{Weg79b} of symmetry class $A$, the model so defined will be referred to simply as the ``Wegner model''. Our interest here is in the regime of strong diagonal disorder, $w_V^2 \gg w_T^2$, on a graph $\mathbb{G}$ with high coordination number. (Since Wegner's original motivation was the option of a $1/N$ expansion, let us stress that we take $N$ to be fixed, possibly as small as $N=1$.)

By the probability law (\ref{eq:problaW}), the ensemble of Hamiltonians (\ref{eq:Wegner}) is invariant under $\mathrm{U}(N)$ basis transformations,
\begin{equation}\label{eq:gaugeT}
    c_{n\alpha}^\dagger \mapsto  \sum c_{n\beta}^\dagger (U_n)_{\beta\alpha} , \quad
    c_{n\alpha} \mapsto \sum (U_n^\dagger)_{\alpha\beta}\, c_{n\beta} \,,
\end{equation}
of the orbital spaces $\mathbb{C}^N$. This property of local $\mathrm{U}(N)$-invariance, while not shared by the standard Anderson model, is expected to emerge \cite{PS82} under renormalization on large length scales. It has the attractive feature of facilitating the introduction of matrix variables.

In an effort to be as pedagogical as possible, we focus on the case of $N = 1$ in the rest of this section. Thus the orbital index disappears for now, and the Hilbert space is $\ell^2(\mathbb{G})$. The case $N \geq 2$ will be taken up in Sect.\ \ref{sect:4.D}.

\subsection{Fourier-Laplace transform}\label{sect:3.A}

The standard technique used in the field-theoretical approach to Anderson localization (actually, delocalization) is the so-called Hubbard-Stratonovich transformation followed by a saddle-manifold approximation, also known as the self-consistent Born approximation, which ultimately leads to a nonlinear $\sigma$-model \cite{EversMirlin}. Now in the present paper we are interested in Anderson transitions at strong coupling (i.e., for large disorder or a small local density of states). In that situation, the Hubbard-Stratonovich transformation is a dead end. The reason is that the Hubbard-Stratonovich field, denoted by $\widetilde{Q}$, fluctuates \emph{very strongly} along the transverse (or non-symmetry) field directions, thereby dooming any kind of controlled saddle approximation. The good field to introduce is a different one, namely the field $Q$ which is dual to $\widetilde{Q}$ by Fourier transformation. Indeed, by the basic principles of Fourier duality, that dual field $Q$ fluctuates \emph{very weakly} (when $\widetilde{Q}$ fluctuates very strongly) along the non-symmetry directions. Thus in our situation of interest, the transverse fluctuations of $Q$ can be controlled, whereas those of $\widetilde{Q}$ are out of control.

The good field $Q$ is introduced by associating with the ensemble of random Schr\"odinger operators $h$ its Fourier-Laplace transform or characteristic functional. This is done as follows. As before, let $n$ denote the sites of our lattice or graph $\mathbb{G}$, and consider two complex scalar fields $n \mapsto u_n$ and $n \mapsto v_n$. Given a Hamiltonian matrix $h_{n n^\prime}$ (for $N=1$) and a real energy parameter $E$, the Hermitian scalar product $\langle \cdot , \cdot \rangle$ of the single-particle Hilbert space $\ell^2(\mathbb{G})$ determines a ``retarded'' inner product ($\varepsilon > 0$),
\begin{equation}
    \langle u , (h - E - \mathrm{i}\varepsilon) u \rangle  \equiv \sum \bar{u}_n (h - E - \mathrm{i}\varepsilon)_{n n^\prime} u_{n^\prime} \,,
\end{equation}
and similarly an ``advanced'' one, $\langle v , (h-E+\mathrm{i}\varepsilon) v \rangle$. The said Fourier-Laplace transform then is a functional, $W$, of the two fields $u$ and $v$:
\begin{align}\label{eq:def-W}
    W[u,v] &= \mathbb{E}\, \Big\{ \mathrm{Det}\big( (h-E)^2 + \varepsilon^2 \big) \cr &\times \mathrm{e}^{-\mathrm{i} \langle u , (h-E-\mathrm{i}\varepsilon) u \rangle}  \mathrm{e}^{+\mathrm{i} \langle v , (h-E+\mathrm{i}\varepsilon) v \rangle } \Big\},
\end{align}
where $\mathbb{E}$ denotes the average over the randomness in $h$.

$W$ contains a wealth of information about the random Schr\"odinger operator $h$ and the physical observables associated to it. Indeed, by integrating $W$ with Lebesgue measure $\prod_n d^2 u_n \, d^2 v_n$ against suitable polynomials in $u$ and $v$ one obtains disorder averages of products of retarded and advanced Green's functions. (The regularization parameter $\varepsilon > 0$ ensures the convergence of these integrals.) The latter can be used to compute transport, spectral and wavefunction observables.

For later reference, we note that $W$ has a global $\mathrm{U}(1)$ symmetry
\begin{equation}\label{eq:globalU1}
    u_n \mapsto \mathrm{e}^{\mathrm{i}\phi} u_n \,, \quad  v_n \mapsto \mathrm{e}^{-\mathrm{i}\phi} v_n \,,
\end{equation}
rotating the phase in the \emph{opposite} way between the retarded and advanced sectors. (Of course, there is also the $\mathrm{U}(1)$ symmetry that transforms $u$ and $v$ by a common phase factor. The latter plays no role here.) Such a symmetry may be broken spontaneously in high enough space dimension. This possibility and its interpretation in the present context will be a major theme of our paper.

\subsection{Switching to matrix variables}\label{sect:3.B}

For the expository account of the present paper, we omit the determinant in Eq.\ (\ref{eq:def-W}). This is done for pedagogical purposes only -- the determinant will of course be re-instated in due course \cite{AZ24}, by its representation as a Gaussian integral over Grassmann variables. (The resulting fermion-fermion sector turns out to be crucial in the final analysis.) Speaking the language of the replica trick, we first deal with the one-replica theory.

In the absence of the determinant factor, the disorder expectation $\mathbb{E}$ is easy to carry out. For the Wegner model, one gets $W = \mathrm{e}^{-S[Q]}$,
\begin{align}\label{eq:SofQ}
    S[Q] &=  w_T^2 \sum_{\langle n n^\prime \rangle} \mathrm{Tr} \, Q_n Q_{n^\prime} \cr &+ \sum_n \mathrm{Tr} \Big( {\textstyle{\frac{1}{2}}} w_V^2 Q_n^2 + \mathrm{i}E Q_n + \varepsilon \sigma_3 Q_n \Big) ,
\end{align}
where $Q_n$ is a $(2 \times 2)$-matrix field
\begin{equation}\label{eq:rankone}
    Q_n = \begin{pmatrix} \bar{u}_n u_n&- \bar{u}_n v_n \cr \bar{v}_n u_n &- \bar{v}_n v_n \end{pmatrix}  = \begin{pmatrix} \bar{u}_n \cr \bar{v}_n \end{pmatrix} \otimes (u_n \; v_n) \, \sigma_3
\end{equation}
with the physical dimension of reciprocal energy. The variances $w_V^2$ and $w_T^2$ were introduced in Eq.\ (\ref{eq:problaW}), and the sum $\sum_{\langle n n^\prime \rangle} ... \equiv \frac{1}{2} \sum A_{n n^\prime} ...$ runs over pairs of adjacent sites. A noteworthy feature is that the initial dependence of $W$ on the complex vector fields with components $u_n$ and $v_n$ has been repackaged in terms of the matrix field $Q_n$. The possibility for this vector-to-matrix reformulation is a consequence of the local gauge invariance (\ref{eq:gaugeT}) of the Wegner model. For present use, let us record two properties that result from the formula (\ref{eq:rankone}):
\begin{equation}\label{eq:pseudoHerm}
    Q_n = \sigma_3 Q_n^\dagger \sigma_3
\end{equation}
and, expressing the fact that $Q_n$ has rank one,
\begin{equation}\label{eq:Det=0}
    \mathrm{Det}(Q_n) = 0 .
\end{equation}

We finish here with a remark on symmetry: in the limit of sending $\varepsilon  \to 0$, the functional $S[Q]$ becomes invariant under global pseudo-unitary transformations
\begin{equation}\label{eq:pseudo}
    Q_n \mapsto g Q_n g^{-1} , \quad g^{-1} = \sigma_3 g^\dagger \sigma_3 \in \mathrm{SU}(1,1) .
\end{equation}
Several of the specific and peculiar features of the field theory of Anderson transitions are due to the fact that the symmetry group $\mathrm{SU}(1,1)$ is noncompact.

We observe that the functional $S[Q]$ for $E = 0$ is real-valued and positive. Moreover, by passing to dimensionless variables we see that the system's behavior is controlled by the effective disorder parameter $w_V^2 / w_T^2$. Our interest is in the strong-coupling regime $w_V^2 \gg w_T^2$.

\section{Strong-disorder limit}\label{sect:4}

To prepare the analysis, we expand $Q$ (suppressing the lattice site index $n$) as
\begin{equation}
    Q = x^0 \mathbf{1} - x^1 \,\mathrm{i}\sigma_1 - x^2 \,\mathrm{i} \sigma_2 + x^3 \sigma_3 \,,
\end{equation}
thereby implementing the pseudo-Hermiticity (\ref{eq:pseudoHerm}) in terms of four real scalars $x^\mu$ ($\mu = 0, \ldots, 3$). Constrained by Eq.\ (\ref{eq:Det=0}), these variables lie in the solution set of the quadratic equation
\begin{equation}
    (x^0)^2 + (x^1)^2 + (x^2)^2 - (x^3)^2 = 0 ,
\end{equation}
with $2 x^3 = \mathrm{Tr} \, \sigma_3 Q = |u|^2 + |v|^2 \geq 0$. This equation can be viewed as the on-shell condition in $2+1$ dimensions for a relativistic particle with invariant mass $|x^0|$, momentum components $(x^1, x^2)$ and energy $x^3 \geq |x^0|$.

Expressed in terms of the $x^\mu$, the functional (\ref{eq:SofQ}) reads
\begin{align}\label{eq:action1}
    S &= 2 \sum_n \left( w_V^2 (x_n^0)^2 + \mathrm{i} E x_n^0 + \varepsilon x_n^3 \right) \\ &+ 2 w_T^2 \sum_{\langle n n^\prime \rangle} \left( x_n^0 x_{n^\prime}^0 - x_n^1 x_{n^\prime}^1 - x_n^2 x_{n^\prime}^2  + x_n^3 x_{n^\prime}^3 \right) . \nonumber
\end{align}
According to the mapping $(u_n , v_n) \mapsto Q_n$ by Eq.\ (\ref{eq:rankone}), the integration measure $\prod_n d^2 u_n \, d^2 v_n$ for the original integration variables gets pushed forward to a product measure $\prod_n d\mu(Q_n)$ with factors (site index $n$ omitted)
\begin{equation}\label{eq:inv-meas}
    d\mu(Q) = dx^0 \, \frac{dx^1 dx^2}{x^3}
\end{equation}
for the composite field $Q$. The factor $dx^1 dx^2 / x^3$ is known from special relativity as the Lorentz-invariant measure on the mass shell
\begin{equation}\label{eq:masshell}
    x^3 = \sqrt{(x^0)^2 + (x^1)^2 + (x^2)^2}
\end{equation}
(for any fixed $x^0$). In the next subsection, we will want to express the measure $d\mu(Q)$ in different coordinates. For that purpose, we note that the Minkowski metric
\begin{equation}\label{eq:Mink}
    - (dx^3)^2 + (dx^1)^2 + (dx^2)^2 = - \frac{1}{2} \mathrm{Tr} \, (dQ^\prime)^2 ,
\end{equation}
where $Q^\prime = x^3 \sigma_3 - x^1 \,\mathrm{i}\sigma_1 - x^2 \,\mathrm{i} \sigma_2$ is the traceless part of $Q$, restricts to a Riemannian metric on every mass shell (i.e., for any fixed invariant mass $|x^0|$). That restriction induces the Riemannian volume element
\begin{equation}\label{eq:invtvol}
    \mathrm{dvol}_{x^0}^\prime = |x^0|\, \frac{dx^1 dx^2}{x^3} \,.
\end{equation}
Thus we have the formula
\begin{equation}\label{eq:form-dmQ}
    d\mu(Q) = \frac{dx^0}{|x^0|} \,  \mathrm{dvol}_{x^0}^\prime \,,
\end{equation}
which will be useful later.

An important observation from Eq.\ (\ref{eq:action1}) is that fluctuations of the invariant-mass variables $x_n^0$ are strongly suppressed in the strong-coupling limit $w_V^2 \gg w_T^2$. (In contrast, the corresponding variable in the Hubbard-Stratonovitch field exhibits large fluctuations beyond our control.) To get an idea of the consequences, let us inspect  the action functional $S$ (for $E = 0 = \varepsilon$) upon restriction to the null surface $x_n^0 = 0$ (for all $n$), where the integrand of our $Q$-integral is maximal:
\begin{equation}\label{eq:aniso}
    S \big\vert_{x_n^0  =0} = 2\sum_{\langle n n^\prime \rangle} \mathrm{e}^{t_n + t_{n^\prime} } \big(1 -\cos(\phi_n - \phi_{n^\prime})\big) .
\end{equation}
Here we have introduced the parametrization
\begin{equation}
    x_n^1 + \mathrm{i} x_n^2 = x_n^3 \, \mathrm{e}^{\mathrm{i}\phi_n} , \quad x_n^3 = \mathrm{e}^{t_n} / w_T
\end{equation}
with a real-valued field $t_n$ and an angular field $\phi_n$.

According to Eq.\ (\ref{eq:aniso}) there is a very large anisotropy of field stiffness. Indeed, if the fluctuating $t$-field populates the real axis uniformly (which is in fact what happens due to symmetry restoration in the localized regime) and the space dimension is high enough (so as to offset the likelihood of a cluster of small $\mathrm{e}^{t_n}$), then spatial variations of the $\phi_n$ are suppressed by a large stiffness. This may lead to spontaneous symmetry breaking for the $\phi$-field. At the same time, if the $\phi$-field settles into a symmetry-broken state $\phi_n = \mathrm{const}$ (independent of $n$), then the $t$-field is free to fluctuate with zero energy cost!

These simple observations raise the intriguing question whether, under suitable circumstances (namely, moderately strong disorder and high dimension or high graph connectivity), the compact symmetry $\phi_n \to \phi_n + \delta\phi$, which was highlighted earlier as the global $\mathrm{U}(1)$ symmetry (\ref{eq:globalU1}), might be broken, while the noncompact symmetry $t_n \to t_n + \delta t$ remains unbroken. This exotic possibility would realize a non-standard phase of disordered electronic matter --- which is the subject of our investigations in the present and subsequent papers.

\subsection{Adjusting the parametrization of $Q$}\label{sect:4.A}

We next take into account the weak fluctuations of the massive variables $x_n^0$, with the expectation that these will induce a tiny stiffness for the $t$-field. To begin, we should issue a warning: the straightforward idea of expanding Eq.\ (\ref{eq:SofQ}) directly in the $x_n^0$ does not work. For example, taking $x^1$ and $x^2$ as the independent variables, one encounters the problem that $x^3$ as given in Eq.\ (\ref{eq:masshell}) is non-analytic in $x^0$ at $x^1 = x^2 = 0$.

Let us therefore adjust the parametrization of the matrix $Q$ so as to make the small-$x^0$ expansion more tractable. We have already seen in Eq.\ (\ref{eq:pseudo}) that the symmetry group $\mathrm{SU}(1,1)$ acts on $Q$ by pseudo-unitary transformations $Q \mapsto g Q g^{-1}$. By virtue of this action, we can set
\begin{equation}\label{eq:orbfac}
    Q_n = g_n^{\vphantom{-1}} q_n^{\vphantom{-1}} g_n^{-1}, \quad g_n \in \mathrm{SU}(1,1) ,
\end{equation}
thereby organizing the field target space into orbits of the symmetry group.

To implement the factorization (\ref{eq:orbfac}) concretely, we need to fix a choice of base point $q$ for each orbit. Whenever $x^0 \not= 0$, there is a natural choice of base point, namely the ``rest frame'' $x^1 = x^2 = 0$ and $x^3 = |x^0|$ for our relativistic particle with invariant mass $|x^0|$. On the other hand, in the especially important case of the massless orbit (or null-orbit), i.e.\ for $x^0 = 0$, no distinguished choice of base point exists (as there is no such thing as a rest frame for a massless relativistic particle). Another aspect of the same issue is that $Q$ cannot be made diagonal on the null-orbit. Indeed, the matrix
\begin{equation}\label{eq:nilpot}
    Q \big\vert_{x^0 = 0} = \begin{pmatrix} |x| &\mathrm{i} \bar{x} \cr \mathrm{i} x &- |x| \end{pmatrix} , \quad - x = x^1 + \mathrm{i} x^2 ,
\end{equation}
is nilpotent (i.e.\ squares to zero) and has the Jordan-normal form $\begin{pmatrix} 0 &|x| \cr 0 & 0 \end{pmatrix}$.

This characteristic feature of the orbit structure entail an important message, as follows. We might be inclined to ignore the complications due to the null-orbit and simply take $q$ to be diagonal ($n$ again omitted):
\begin{equation}\label{eq:diag-q}
    Q = g \begin{pmatrix} x^0 + |x^0| &0 \cr 0 &x^0 - |x^0| \end{pmatrix} g^{-1} ,
\end{equation}
in keeping with standard practice in the field theory approach to Anderson localization. If so, we would run into a serious difficulty: the special value $x^0 = 0$ would no longer correspond to the null-orbit $x^3 = |x^1 + \mathrm{i} x^2|$ but would give no more than the zero matrix. At the same time, the null-orbit would be (mis-)represented as a coordinate singularity; indeed, the integration measure (\ref{eq:form-dmQ}) in the parametrization (\ref{eq:diag-q}) is non-analytic in $x^0$:
\begin{equation}
    d\mu(Q) = |x^0|\, dx^0 dg_{G/K} .
\end{equation}
Here $dg_{G/K}$ is the $G$-invariant measure on the standard hyperboloid $G/K = \mathrm{SU}(1,1) / \mathrm{U}(1) = \mathrm{H}^2$.

This brings us to a key technical message of the present paper: to make progress with our goal of eliminating the massive degrees of freedom transverse to the null-orbit, we require the orbit parametrization, Eq.\ (\ref{eq:orbfac}), to be smooth in the variable $x^0$. For that reason, we must abandon the diagonal choice for $q$.

Guided by these observations, and in order to accommodate the nilpotency of $Q\big\vert_{x^0 = 0}$ in Eq.\ (\ref{eq:nilpot}), we take $q_n \equiv q$ to be given by
\begin{align}\label{eq:q-green}
    &x^0(q) = \nu m , \quad x^3(q) = \frac{\nu}{2} (1 + m^2), \cr &(x^1 + \mathrm{i} x^2)(q) = \frac{\mathrm{i}\nu}{2} (1-m^2) \,\mathrm{e}^{\mathrm{i}\vartheta} .
\end{align}
 The variable $m \in \mathbb{R}$ is a dimensionless variant of the fluctuating invariant mass $x^0$. (Note that the null-orbit is still at $m = 0$.) The dimensionful parameter $\nu$ sets the unit on the reciprocal energy scale; without loss, we presently set it to unity and redefine all energies $(E, \varepsilon, w_V, w_T)$ as dimensionless multiples of $\nu^{-1}$. (The parameter $\nu$ has the physical meaning of a density of states per site and energy, which is a scale to be dynamically generated in the field theory.) Fixed at some arbitrary value -- recall once more that no preferred frame exists for a massless relativistic particle -- the angular parameter $\vartheta$ is a peculiar feature forced by the central role of the null-orbit $m=0$. Note that any fixed value of $\vartheta$ breaks the $\mathrm{U}(1)$ symmetry (\ref{eq:globalU1}). In keeping with our efforts to be pedagogical, we here set $\vartheta = 0$.

\begin{figure}
    \centering
    \includegraphics[width=8cm]{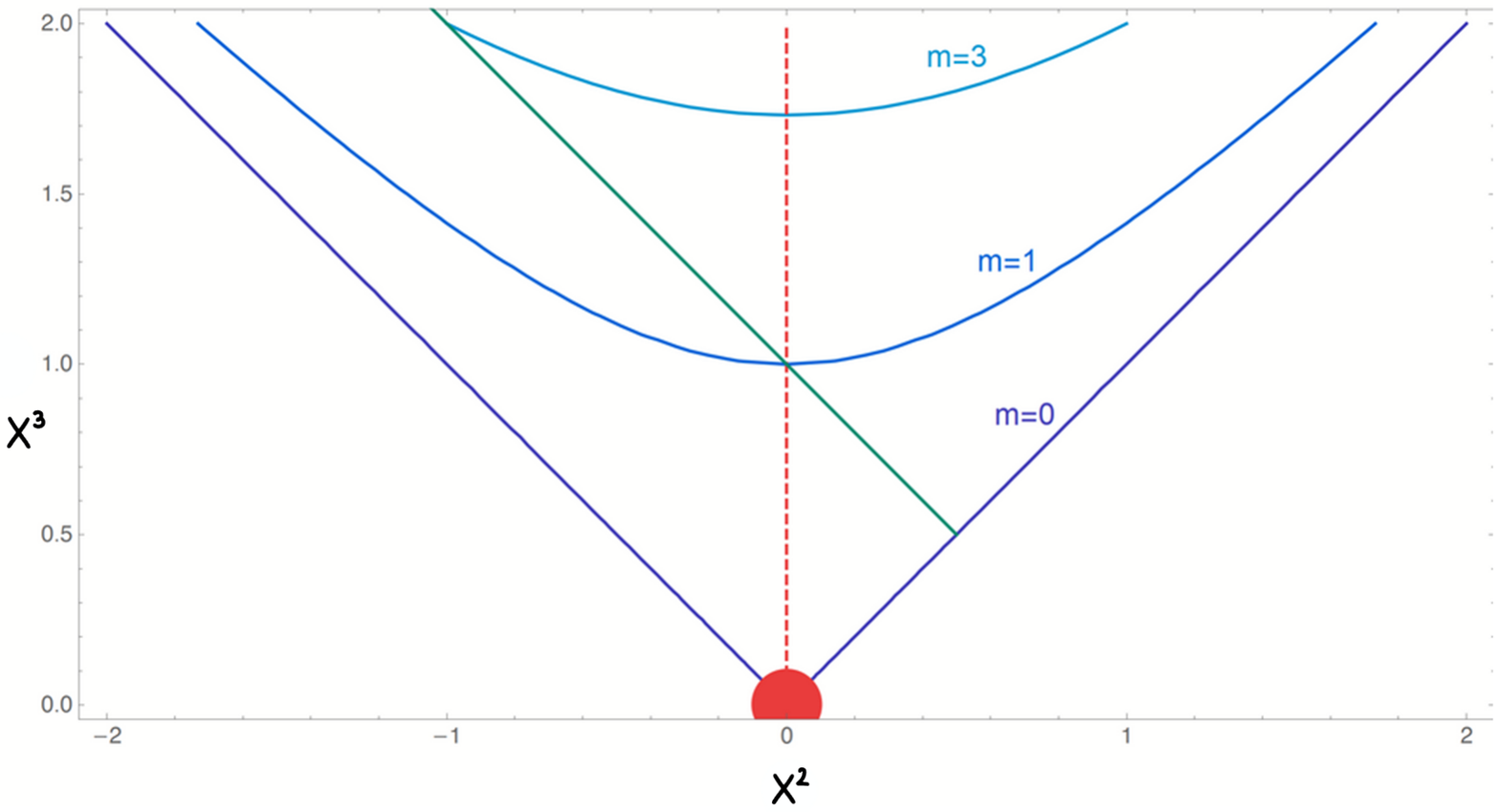}
    \caption{Illustration of $q$ (for $\vartheta = 0$) as a function of $m$ with values in the $x^2 x^3$ plane ($x^1 = 0$). Given by Eq.\ (\ref{eq:q-green}), the trajectory of $q(m)$ is the green line (actually, a double cover thereof, as $m$ can be positive or negative). We note that the green trajectory avoids the cone singularity at $x^2 = x^3 = 0$ (fat red dot). In contrast, the parametrization of $q$ as a diagonal matrix as in Eq.\ (\ref{eq:diag-q}) runs right into the cone singularity (dashed red line). To avoid that non-analyticity, the green line must break the $\mathrm{U}(1)$ symmetry (\ref{eq:globalU1}).} \label{fig:F2}
\end{figure}

The path of  $m \mapsto q(m)$ is shown in Fig.\ \ref{fig:F2}.

\subsection{Elimination of the massive variables}\label{sect:4.B}

To make the following calculation (and its later generalizations with supersymmetry) as transparent as possible, we now carry out a unitary basis transformation $U$ for the advanced-retarded space, $\mathbb{C}_{\rm ar}^2\,$, so that
\begin{equation}
    \sigma_3 \to \sigma_2 \,, \quad \sigma_2 \to \sigma_1 \,, \quad \sigma_1 \to \sigma_3
\end{equation}
or, more succinctly, $U \sigma_j U^{-1} = \sigma_{j-1}$ (where $0 \equiv 3$). We note that the pseudo-Hermiticity (\ref{eq:pseudoHerm}) of $Q$ changes to
\begin{equation}
    Q = \sigma_2 Q^\dagger \sigma_2 \,,
\end{equation}
and the symmetry group $\mathrm{SU}(1,1)$ gets converted to the isomorphic group $\mathrm{Sp}(2,\mathbb{R}) \cong \mathrm{SL}(2, \mathbb{R})$ with real Lie algebra spanned by the real matrices $\sigma_1 , \mathrm{i} \sigma_2, \sigma_3$. A major benefit from this basis change is that $q$ (or even better, the Lie-algebra valued matrix $\mathrm{i} q$) takes the simple form
\begin{equation}\label{eq:iq}
    \mathrm{i} q = \begin{pmatrix} \mathrm{i}m &1 \cr -m^2 &\mathrm{i}m \end{pmatrix} \in \mathfrak{u}(1) \oplus \mathfrak{sl}(2,\mathbb{R})
\end{equation}
(recall the choice $\vartheta = 0$), or with $(\sigma_1 \pm \mathrm{i} \sigma_2) / 2 \equiv  \sigma_\pm \,$,
\begin{equation}
    q = m \cdot 1 - \mathrm{i} \sigma_+ + \mathrm{i} m^2 \sigma_- \,.
\end{equation}
The matrix $Q$ is still given by Eq.\ (\ref{eq:orbfac}) but now with real transformations $g$ from $\mathrm{SL}(2, \mathbb{R})$. Inserting $Q_n = g_n q_n g_n^{-1}$ into Eq.\ (\ref{eq:SofQ}), we obtain the following exact reformulation of the action functional $S$ (for $\varepsilon \to 0$):
\begin{align}\label{eq:action-2}
    S &= 2 \sum_n \left( w_V^2 m_n^2 + \mathrm{i} E m_n \right) + w_T^2 \sum\nolimits_{\langle n n^\prime\rangle}  \\
    &\Big\{ (m_n + m_{n^\prime})^2 - \mathrm{Tr}\, Q_n^+ Q_{n^\prime}^+ - m_n^2 m_{n^\prime}^2 \mathrm{Tr}\, Q_n^- Q_{n^\prime}^- \cr
    &+ m_n^2 (\mathrm{Tr} \, Q_n^- Q_{n^\prime}^+ - 1) + m_{n^\prime}^2 (\mathrm{Tr}\, Q_n^+ Q_{n^\prime}^- - 1)\Big\} , \nonumber
\end{align}
where we have abbreviated
\begin{equation}
    g_n \, \sigma_{\pm} \, g_n^{-1} \equiv Q_n^\pm \,.
\end{equation}
We observe that the expression (\ref{eq:action-2}) for $S$ depends analytically (and uniformly so) on the variables $m_n$. Moreover, the integration measure is analytic in $m_n$ as long as good orbit coordinates are chosen. For example, taking
\begin{equation}\label{eq:def-st}
    g = \exp \left( s \sigma_1 / 2 \right) \exp \left( t \sigma_3 / 2 \right) \in \mathrm{SL}(2,\mathbb{R}) ,
\end{equation}
one has the metric tensor
\begin{equation}
    - \frac{1}{2} \mathrm{Tr} \, (dQ^\prime)^2 =
    m^2 dt^2 + (\mathrm{e}^t + m^2 \mathrm{e}^{-t})^2 ds^2 .
\end{equation}
Formula (\ref{eq:form-dmQ}) then gives the integration measure as
\begin{equation}\label{eq:dmuQ}
    d\mu(Q) = (\mathrm{e}^t + m^2 \mathrm{e}^{-t}) \, dm \, ds \, dt .
\end{equation}
At this point, we reiterate the message that, if we had made the diagonal choice (\ref{eq:diag-q}) for $q$, the integration measure (and also the action $S$) would have been non-analytic in the massive variables at the cone singularity, thereby obstructing the upcoming elimination process.

We also note that the present calculation would simplify (and could be performed exactly, without any approximation) if horospherical coordinates were used. We refrain from doing so, as these coordinates do not carry over to the supersymmetric setting and therefore do not serve as a helpful guide to the ultimate picture.

We are finally in a position to eliminate the weakly fluctuating variables $m_n$. To that end, we assume the strong-disorder regime where $w_V$ is the largest energy scale of the system, and we expand in the small ratio $w_T^2 / w_V^2 \ll 1$. At the same time, we assume that our graph $\mathbb{G}$ has a high enough connectivity to force collective-field (if not mean-field) behavior. The latter is to say that all the $\mathrm{Tr}\, Q Q^\prime$ terms in (\ref{eq:action-2}), in particular
\begin{equation}
    \mathrm{Tr}\, Q_n^- Q_{n^\prime}^- \,, \quad \mathrm{Tr} \, Q_n^- Q_{n^\prime}^+ - 1 \,, \quad \mathrm{Tr}\, Q_n^+ Q_{n^\prime}^- - 1
\end{equation}
are typically small compared to unity. (They might be replaced by gradients in a controlled continuum limit.)

Under the specified conditions, the variables $m_n$ can be integrated out in an approximation known as the one-loop order of perturbation theory. Its consummate effect is to replace the $m_n^2$ by their expectation w.r.t.\ the (normalized) distribution with function $\mathrm{e}^{-2(w_V^2 m_n^2 + \mathrm{i} E m_n)}$:
\begin{equation}\label{eq:m2av}
    m_n^2 \longrightarrow \frac{1}{4w_V^2} \left( 1 - \frac{E^2}{w_V^2} \right) \equiv \langle m^2 \rangle .
\end{equation}
Neglecting the $m$-dependence of the integration measure (which gives corrections of higher loop order), we then arrive at the following effective action:
\begin{align}\label{eq:act-eff}
    S_{\rm eff} = w_T^2 \sum\nolimits_{\langle n n^\prime \rangle} &\mathrm{Tr} \Big(
    \langle m^2 \rangle (Q_n^+ Q_{n^\prime}^- +  Q_n^- Q_{n^\prime}^+) \cr
    &- Q_n^+ Q_{n^\prime}^+ - \langle m^2 \rangle^2 \, Q_n^- Q_{n^\prime}^- \Big) .
\end{align}
It is worth noting that the symmetry under global transformations $g_n \mapsto g g_n$ for $g \in \mathrm{SL}(2,\mathbb{R}) \cong \mathrm{SU}(1,1)$ remains fully intact after the elimination of the $m_n$.


\subsection{Nonlinear $\sigma$-model?}\label{sect:4.C}

To get an explicit view of $S_{\rm eff}$ and compare with the formula (\ref{eq:aniso}), we display its parts in polar coordinates,
\begin{equation}
    g_n = \exp\left( \mathrm{i}\phi_n \sigma_2/2 \right) \exp \left( t_n \sigma_3 / 2 \right) .
\end{equation}
Inserting this expression into $Q_n^\pm = g_n \sigma_\pm g_n^{-1}$ we obtain
\begin{align}
    - \mathrm{Tr}\, Q_n^+ Q_{n^\prime}^+ = \mathrm{e}^{t_n + t_{n^\prime}} \left( 1 - \cos(\phi_n - \phi_{n^\prime}) \right) / 2 .
\end{align}
This term, the only survivor in the limit of $\langle m^2 \rangle \to 0$, agrees with what was shown in Eq.\ (\ref{eq:aniso}).
Now as a result of the small $m$-fluctuations, another term has appeared:
\begin{align}
    &\mathrm{Tr} \left( Q_n^+ Q_{n^\prime}^- + Q_n^- Q_{n^\prime}^+ \right) = \cr
    &= \cosh(t_n - t_{n^\prime}) \left( 1 + \cos(\phi_n - \phi_{n^\prime}) \right) .
\end{align}
It equips the $t$-field with a stiffness that is non-vanishing but much reduced by the smallness of the multiplying factor $\langle m^2 \rangle$, cf.\ Eq.\ (\ref{eq:act-eff}). According to (\ref{eq:m2av}) that factor goes to zero in the strong-disorder limit $w_V^2 \to \infty$ and also when the energy $E$ approaches the band edge ($|E| = w_V$) of the disorder-broadened band. The final term,
\begin{align}
    - \mathrm{Tr}\, Q_n^- Q_{n^\prime}^- = \mathrm{e}^{- t_n - t_{n^\prime}} \left( 1 - \cos(\phi_n - \phi_{n^\prime}) \right) / 2,
\end{align}
is doubly suppressed by its coefficient $\langle m^2 \rangle^2$ in the effective action (\ref{eq:act-eff}); we have kept it only for completeness and in order to  a facilitate a remark made below.

The take-away message here is that we have a strong imbalance of couplings: the stiffness of the $t$-field is much smaller than that of the $\phi$-field. Now our attentive readers might raise the following objection. We could make a shift $t_n \to t_n + \frac{1}{2} \ln \langle m^2 \rangle$ of torus variables, taking the effective action to the more symmetric form
\begin{equation}
    S_{\rm eff}^\prime = w_T^2 \langle m^2 \rangle \sum_{\langle n n^\prime \rangle} \mathrm{Tr}\, (Q_n^+ - Q_n^-) (Q_{n^\prime}^- - Q_{n^\prime}^+) ,
\end{equation}
which we recognize as that of the usual lattice nonlinear $\sigma$-model with hyperbolic target space $\mathrm{H}^2$. So, why the big noise about an imbalance of couplings? The answer is yes, one may perform such a shift. Yet, the objection is naught because the possibility of such a shift is a misleading feature of the one-replica theory. Indeed, leaving warm-ups behind, we must think ahead to the final formulation of the theory, which includes a fermion-fermion sector built from Grassmann variables and, there, such a rescaling of couplings is \emph{not available}. (The isometry group of the fermion-fermion sector is a group of unitary transformations that cannot change the scale without changing the field compactification radius.)

Let us compound this with a follow-up remark. One might think that knowledge of the symmetry alone suffices to identify the nonlinear $\sigma$-model uniquely as the proper low-energy effective field theory. That is not quite so. The correct statement is that one needs to know the symmetry in conjunction with the pattern of symmetry \emph{breaking}. In the case of our Anderson localization problem, the latter is not a given. Indeed, as should be clear from the above, the pattern of symmetry breaking at strong coupling is complicated and modified by the fact that our symmetry group $G = \mathrm{SU}(1,1)$ features different $G$-orbit types, and in the strong-disorder limit the field resides dominantly on an atypical $G$-orbit, the null-orbit (or ``light cone'' for a massless relativistic particle).


Thus, returning to the main thread of this subsection, we insist that the strong imbalance of couplings (large $\phi$-stiffness versus small $t$-stiffness) is not fake but a true effect. In fact, as hinted earlier, it is the seed for our scenario of partial symmetry breaking. Jumping ahead to a follow-up paper \cite{AZ24}, let us anticipate that the full effective field theory (in continuum approximation for a system in $d$ space dimensions) can be obtained from Eq.\ (\ref{eq:act-eff}) by a few obvious substitutions:
\begin{equation}\label{eq:S-cont}
   S_{\rm cont} = \int d^d x \left( \lambda_\tau \nabla Q^+ \cdot \nabla Q^- -\lambda_\sigma (\nabla Q^+)^2 \right) .
\end{equation}
Here $Q^\pm$ are $4 \times 4$ supermatrix fields, and $\mathrm{STr}$ denotes the supertrace. The couplings are given by
\begin{equation}\label{eq:coupl}
    \lambda_\sigma = a^{d-2} w_T^2 , \quad \lambda_\tau / \lambda_\sigma = 2 \langle m^2 \rangle \ll 1 ,
\end{equation}
with a length scale $a$ that accounts for the passage from the lattice to the continuum.

Let us then repeat the key message of the present paper: taking the strong-coupling limit seriously and adjusting the field-theory technique accordingly, one discovers that the usual description by a nonlinear $\sigma$-model becomes invalid and the true effective field theory (for strong disorder and/or small local density of states) is another one with the same global symmetry but \underline{two} independent couplings --
a small stiffness $\lambda_\tau$ for the ``light-like'' fluctuations (as measured by $\mathrm{Tr}\, \nabla Q^+ \nabla Q^-$) and a much larger stiffness $\lambda_\sigma$ for the ``space-like'' fluctuations ($\mathrm{Tr}\, \nabla Q^+ \nabla Q^+$).

\subsection{Extension to $N > 1$}\label{sect:4.D}

We now go through the changes that occur when the number $N$ of orbitals is raised to $N \geq 2$. The outcome will be very much the same as for $N=1$, except that the expressions for the bare values of the field stiffness constants $\lambda_\sigma$ and $\lambda_\tau$ get modified by $N$ entering as a further parameter. The changes come about mostly because the formulas for the integration measure and the parametrization of $Q$ need to be adapted slightly.

Assuming the probability law (\ref{eq:problaW}), we quickly arrive at Eq.\ (\ref{eq:SofQ}) for the action functional $S$, albeit with $Q$ now a sum over orbitals:
\begin{equation}
    Q(u,v) = \sum_{\alpha=1}^N \begin{pmatrix} \bar{u}_{\alpha} u_{\alpha}&- \bar{u}_{\alpha} v_{\alpha} \cr \bar{v}_{\alpha} u_{\alpha} &- \bar{v}_{\alpha} v_{\alpha} \end{pmatrix} .
\end{equation}
(Here the site index $n$ has been omitted, and we have returned to the initial basis for the advanced-retarded space $\mathbb{C}^2$.) The space of such matrices $Q$ has real dimension four. We observe that $\mathrm{Det}(\sigma_3 Q) \geq 0$ since $Q \sigma_3$ is generically a positive Hermitian matrix by the Cauchy-Schwarz inequality. Equation (\ref{eq:pseudoHerm}) holds as before. The change of integration variables from pairs of $N$-component complex vectors $\{ u_{n\alpha}\,, v_{n\alpha} \}_{\alpha = 1, \ldots, N}$ to $(2 \times 2)$-matrices $Q_n$ is now effected by the formula \cite{BEKYZ07}
\begin{align}\label{eq:ChangeVar}
    &\int D\mu(Q) \, \mathrm{Det}^N(Q \sigma_3) \, F(Q) \cr &= \int_{\mathbb{C}^N} d^{2N} u \int_{\mathbb{C}^N} d^{2N} v  \, F\big( Q(u,v) ) .
\end{align}
The $Q$-integral here runs over all matrices $Q \sigma_3 > 0$, and $D\mu(Q)$ is an integration measure invariant under all transformations
\begin{equation}
    Q \sigma_3 \to g \, Q \sigma_3 \, g^\dagger , \quad g \in \mathrm{GL}(2,\mathbb{C}) .
\end{equation}
Note that the normalization constant of $D\mu(Q)$ depends on $N$. It is conveniently fixed by evaluating both sides of Eq.\ (\ref{eq:ChangeVar}) for the special function $F(Q) = \mathrm{e}^{-\mathrm{Tr}\, \sigma_3 Q}$, which yields the normalization condition
\begin{equation}
    \int D\mu(Q) \, \mathrm{Det}^N(\sigma_3 Q) \, \mathrm{e}^{- \mathrm{Tr}\, \sigma_3 Q} = \pi^{2N}  .
\end{equation}
We note that elements $g$ of the symmetry group  $\mathrm{SU}(1,1)$ still act on $Q$ by conjugation; cf.\ Eq.\ (\ref{eq:orbfac}).

At this point, we again make the change of basis taking $\sigma_j$ to $\sigma_{j-1}$. Using the action (\ref{eq:orbfac}) by $\mathrm{SU}(1,1) \cong \mathrm{SL}(2,\mathbb{R})$, we adapt the parametrization (\ref{eq:iq}) to
\begin{equation}\label{eq:iq-p}
    \mathrm{i} q = \begin{pmatrix} \mathrm{i}m &1 \cr -m^2-p &\mathrm{i}m \end{pmatrix}
\end{equation}
with $p$ real and positive. We then have the formulas
\begin{equation}
    \mathrm{Det}(\sigma_2 Q) = \mathrm{Det}(\mathrm{i} q) = p ,
\end{equation}
and
\begin{equation}
    \mathrm{Tr} \, Q^2 = 4 m^2 + 2p .
\end{equation}
A global coordinate system for the space of matrices $Q$ is provided by the variables $m, p$ augmented by the variables $s, t$ defined by Eq.\ (\ref{eq:def-st}). It is straightforward to show that the coordinate expression for $D\mu(Q)$ is
\begin{equation}
    D\mu(Q) = c_N \big( \mathrm{e}^t + m^2 \mathrm{e}^{-t} \big) \, \frac{dp}{p^2} \, dm \, ds \, dt
\end{equation}
with normalization constant
\begin{equation}
    c_N = \frac{\rm const}{\Gamma(N) \Gamma(N-1)} .
\end{equation}
Note that $c_N$ vanishes for $N = 1$. This is one indication that a uniform treatment for all $N \geq 1$ is impossible.

With all that information in hand, we can repeat the process of elimination of the massive degrees of freedom $m_n$ and $p_n$ of $Q_n$.
In view of Eq.\ (\ref{eq:iq-p}) compared with Eq.\ (\ref{eq:iq}), all squares $m_n^2$ in the expression (\ref{eq:action-2}) for $S$ get shifted to $m_n^2 + p_n$. The weakly fluctuating variables $m_n^2$ are handled exactly as before. The weak fluctuations of the variables $p_n$ are controlled by the factor $\mathrm{e}^{- w_V^2 \sum_n p_n}$ from $\mathrm{e}^{-S}$ in conjunction with the factors from $\prod_n D\mu(Q_n)\, \mathrm{Det}^N(\sigma_2 Q_n)$. Taking these together, the $p_n$ are eliminated by integration against the measure
\begin{equation}\label{eq:p-meas}
    \mathcal{N}^{-1} \prod_n \frac{dp_n}{p_n^2} \, p_n^N \, \mathrm{e}^{- w_V^2 p_n} .
\end{equation}
The outcome is that $\langle m^2 \rangle$ in Eq.\ (\ref{eq:coupl}) gets replaced by $\langle m^2 + p \rangle$ with
\begin{equation}
    \langle p \rangle = \frac{N-1}{w_V^2}
\end{equation}
as computed from (\ref{eq:p-meas}). Thus our final expressions for the unrenormalized couplings of the $N$-orbital Wegner model are
\begin{align}\label{eq:coupfin}
    \lambda_\sigma &= a^{d-2} w_T^2 \,, \\
    \lambda_\tau &= a^{d-2} \frac{w_T^2}{2w_V^2} \left( 4N - 3 - \frac{E^2}{w_V^2} \right) . \nonumber
\end{align}
We see that in the band center ($E = 0$) the inequality $\lambda_\tau \ll \lambda_\sigma$ holds  in the strong-disorder limit $w_V^2 \gg N w_T^2$.

\section{RG flow diagram: predictions}\label{sect:5}

\begin{figure}
    \centering
    \includegraphics[width=8cm]{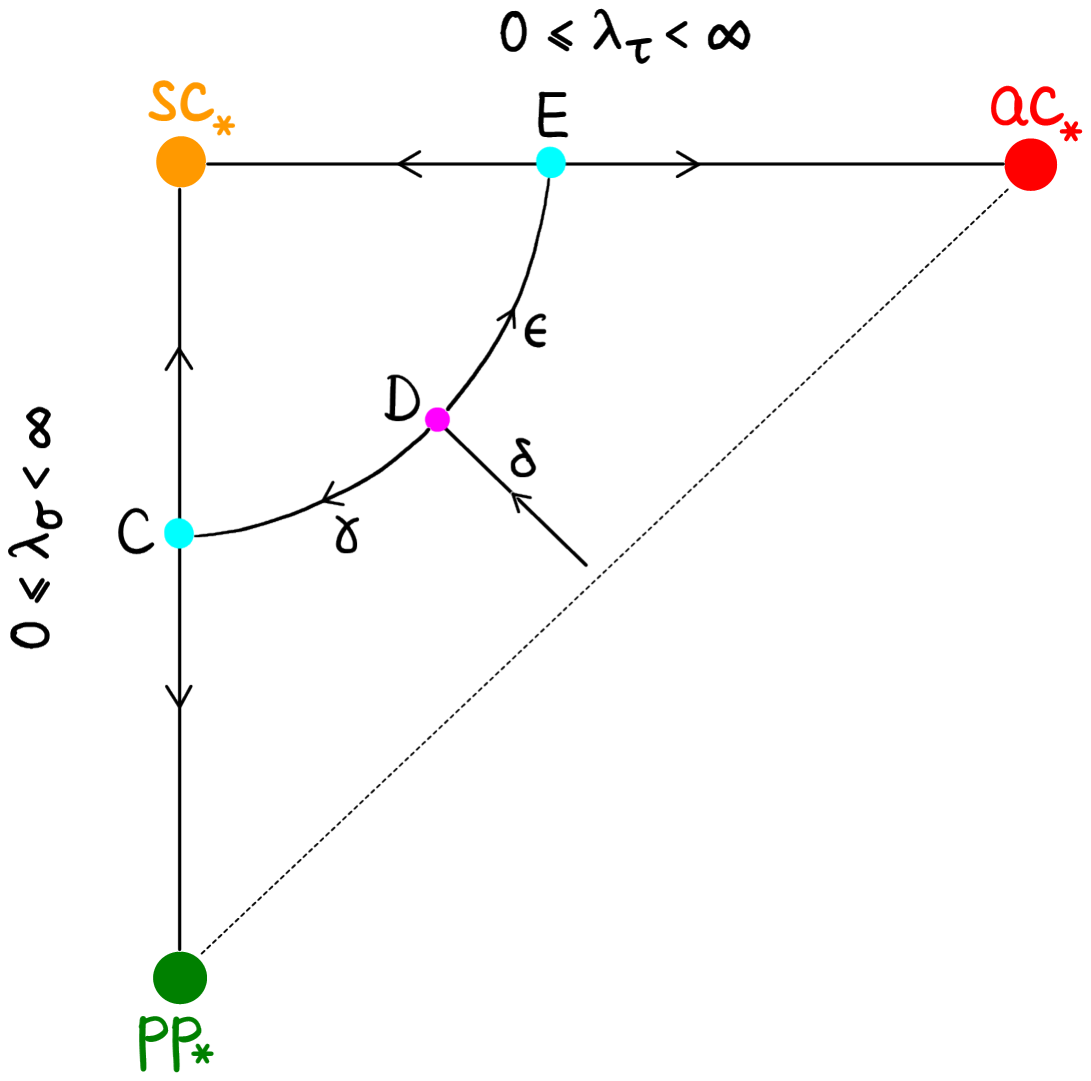}
    \caption{Skeleton of the  conjectured renormalization group (RG) flow diagram (Fig.\ \ref{fig:F1}). The diagram features three attractive RG-fixed points: ${\rm pp}_\ast$, ${\rm ac}_\ast$, and ${\rm sc}_\ast$, with corresponding basins of attraction. The phase boundaries between them are pp/ac $\equiv \delta$, ac/sc $\equiv \epsilon$, and sc/pp $\equiv \gamma$. These phase boundaries meet at a multi-critical point, $D$. The RG-flow along the boundary lines $\gamma$ and $\epsilon$ terminates at unstable RG-fixed points $C$ resp.\ $E$. The flow along $\delta$ becomes marginal near $D$.} \label{fig:F3}
\end{figure}

To conclude the present paper, we continue the discussion of the proposed RG flow diagram (Fig.\ \ref{fig:F1}). Recall that its central feature is the attractive RG-fixed point ${\rm sc}_\ast$ (whose existence still needs to be shown) at $\lambda_\tau = 0$ and $\lambda_\sigma = \infty$. Taking ${\rm sc}_\ast$ and the stable flow into it for granted, the gross features of the RG flow diagram follow (Fig.\ \ref{fig:F3}). The open basin of attraction of ${\rm sc}_\ast$ is delineated by two phase boundaries, abbreviated here as ac/sc $\equiv \epsilon$ and sc/pp $\equiv \gamma$. On both $\gamma$ and $\epsilon$ the flow departs from the multi-critical point $D$ and arrives at the unstable RG-fixed points $C$ resp.\ $E$. The flow into $D$ along the phase boundary pp/ac $\equiv \delta$ must be marginal, as it picks up again on the other side and heads towards ${\rm sc}_\ast$.

What are the immediate predictions (independent of all the numerical details that vary with space dimension and universality class) which can be made on the basis of the RG flow diagram (\ref{fig:F3})? A robust prediction is two-parameter scaling with two critical length scales. Indeed, consider approaching the phase boundary $\delta$ from the ac-side. You are then going to observe two divergent length scales: the first (and smaller) one, $\xi_1$, is the exponential $\mathrm{e}^t$ of the time (i.e.\ the increase of cutoff scale $a \to {\rm e}^t a$) for the RG flow to get past the region of critical slowing down near $D$ and head toward $\lambda_\sigma(E) = \infty$; the second (and larger) one, $\xi_2$, is the exponentiated time to get past $E$. The smaller length is to be interpreted as the scale where it becomes discernible that the critical wave function does not localize, while the larger length is the scale over which the wave function develops full ergodicity; cf.\ \cite{SLS2022}.

A similar discussion, with $C$ taking the role of $E$, can be given for the critical behavior on the pp-side of $\delta$. The smaller length is now the scale at which it becomes apparent that the RG flow is headed towards $C$ with $\lambda_\tau(C) = 0$ instead of $E$ with $\lambda_\sigma(E) = \infty$. The larger length is the scale at which the critical RG flow has gotten past $C$ and is rapidly approaching ${\rm pp}_\ast$. To interpret this in physics terms, let us assume for near-critical but still localized wave functions the picture \cite{Lemarie2022} of polymer-shaped clusters with a longitudinal localization length $\xi_\parallel$ large compared to the transversal localization length $\xi_\perp$. Then our smaller length is $\xi_\perp$, and the larger one is $\xi_\parallel$. Here we should note that if the boundary of the pp-phase is approached along $\gamma$, then $\xi_\parallel$ diverges, while $\xi_\perp$ only increases to a universal finite value \cite{Lemarie2022}. The latter happens because $D$ (and the concomitant divergence due to critical slowing down) lies in the RG-past for $\gamma$.

To make these rough predictions quantitative, one needs to compute the RG-beta vector field (``beta'') near the points $C$, $D$, and $E$. While that is a very difficult computation by analytical means (and therefore in 3D is best left to numerical simulations, albeit the high-D case might be tractable by mean-field methods), the qualitative outcome is constrained by the analyticity of beta.

First, consider the situation near $D$. Suppose that the straight line given as the solution set of
\begin{equation}\label{eq:unstable}
    r_D \big( \lambda_\tau - \lambda_\tau(D) \big) = \lambda_\sigma - \lambda_\sigma(D) \equiv \delta\lambda_\sigma
\end{equation}
is tangent to the unstable manifold ($\gamma \cup \epsilon$) at $D$, and
\begin{equation}\label{eq:marginal}
      s_D \big( \lambda_\sigma - \lambda_\sigma(D) \big) = \lambda_\tau - \lambda_\tau(D) \equiv \delta\lambda_\tau
\end{equation}
is tangent to the marginal flow line $\delta$. ($r_D$ and $s_D$ are real numbers which we expect to be positive if the drawing of Fig. \ref{fig:F3} reflects the actual situation.) Then the RG-beta ``function'' near $D$ is the analytic vector field
\begin{align}\label{eq:betaD}
    \beta^{(D)} &= \frac{1}{\nu_D} \big( \delta\lambda_\tau - s_D \, \delta\lambda_\sigma \big) \left(
    \frac{\partial}{\partial\lambda_\tau} + r_D \frac{\partial}{\partial\lambda_\sigma}
    \right) \\ &+ y_D \big( \delta\lambda_\sigma - r_D \, \delta\lambda_\tau \big)^2
    \left( \frac{\partial}{\partial\lambda_\sigma} + s_D \frac{\partial}{\partial\lambda_\tau}
    \right) . \nonumber
\end{align}
(Here we are using the symbol of partial derivative to denote the basis vector fields given by the coordinate functions $\lambda_\sigma$ and $\lambda_\tau$.) This is verified by observing that the vector field (or first-order differential operator) $\beta^{(D)}$ annihilates the function
\begin{equation}
    f_D = \left\vert \delta\lambda_\tau - s_D \, \delta\lambda_\sigma \right\vert^{\nu_D} {\rm e}^{y_D^{-1} (  \delta \lambda_\sigma - r_D \, \delta\lambda_\tau)^{-1}} ,
\end{equation}
and the solution set of the characteristic equation $f_D = 0$ (approaching $\gamma \cup \epsilon$ from the non-sc side) consists of the line (\ref{eq:unstable}) of instability and the line (\ref{eq:marginal}) of marginal flow. By computing the RG flow in the linear approximation given by $\beta^{(D)}$, one finds that $\xi_1$ on approaching $\delta$ exhibits a power-law divergence with exponent $\nu_ 1 = \nu_D$.

Second, consider the RG-fixed point $E$. If the stable manifold $\epsilon$ near $E$ is given by the equation $\lambda_\tau = \lambda_\tau(E) - r_E / \lambda_\sigma$ (for some real number $r_E$), then the leading approximation to the RG-beta vector field at $E$ is
\begin{align}\label{eq:betaE}
    \beta^{(E)} &= \frac{1}{\nu_E} \left( \lambda_\tau - \lambda_\tau(E) + \frac{r_E}{\lambda_\sigma} \right) \frac{\partial}{\partial \lambda_\tau} \cr &+ y_E \left( \lambda_\sigma \frac{\partial}{\partial\lambda_\sigma} + \frac{r_E}{\lambda_\sigma} \, \frac{\partial}{\partial\lambda_\tau} \right).
\end{align}
To verify (\ref{eq:betaE}), we note that $\beta^{(E)}$ annihilates the function
\begin{equation}
    f = \big\vert \lambda_\tau - \lambda_\tau(E) + r_E / \lambda_\sigma \big\vert^{\nu_E} \lambda_\sigma^{-1/y_E} ,
\end{equation}
and the solution set of the characteristic equation $f_E = 0$ consists of the stable manifold given by $\lambda_\tau - \lambda_\sigma(E) + r_E / \lambda_\sigma = 0$ and of the line $\lambda_\sigma^{-1} = 0$ of instability. By tracking the RG flow close to the separatrix $\delta \cup \epsilon$, we see that the critical length $\xi_2$ diverges near $\gamma$ as a power law with exponent $\nu_2 = \nu_D + \nu_E$.
For example, if $\nu_D = \nu_E = 1/2$ then \cite{SLS2022}
\begin{equation}
    \nu_1 = \nu_D = 1/2, \quad \nu_2 = \nu_D + \nu_E = 1 .
\end{equation}
The numbers $y_D, y_E$ in Eqs.\ (\ref{eq:betaD}, \ref{eq:betaE}) are irrelevant scaling dimensions.

The RG-fixed point $C$ instead of $E$ is handled in the same way: we just flip $\lambda_\sigma \leftrightarrow \lambda_\tau$ and take the reciprocal ($\lambda \to \lambda^{-1}$) to exchange zero and infinity.

\section{Conclusions}\label{sect:6}

The quantum critical phenomenon of the Anderson transition in high dimension ($d \geq 3$) takes place in a regime of strong disorder. Getting an analytical grip on it poses a challenge for field theory and the renormalization group. To complicate matters, the traditional approach via Hubbard-Stratonovich (HS) transformation fails for the strong-coupling situation at hand, as the requisite saddle-point approximation (a.k.a.\ SCBA) is out of control. Fortunately, there exists a viable alternative: the good field to introduce at strong coupling is the bosonization field dual to the HS field. The objective of the present paper was to describe the main steps of a controlled derivation of the bosonized field theory.

In an effort to render our expository account as pedagogical as possible, we illustrated the field-theory derivation at the example of one replica (in place of the zero-replica limit). Needless to say, that theory is deficient and gives the wrong physics. Nevertheless, our readers should rest assured that the extension to zero replicas (actually, the ``supersymmetric'' version) is more or less immediate and will be provided in a forthcoming paper. There is only one substantial change to be implemented: bosonization has to be extended to ``superbosonization'', which is readily available for the Wegner model with $N \ge 2$ orbitals; cf.\ \cite{BEKYZ07} and \cite{LSZ08}, Sect.\ 1.3.

The emerging field theory at strong coupling has two parameters: a small stiffness $\lambda_\tau$ for the field fluctuations tangential to a light-like (or metric-degenerate) submanifold of the target space, and a much larger stiffness $\lambda_\sigma$ for the field fluctuations transverse to it. (The usual nonlinear $\sigma$-model is recovered by setting $\lambda_\sigma = \lambda_\tau$.) Satisfying basic consistency requirements, the theory for $\lambda_\sigma \gg \lambda_\tau$ is still invariant under the Killing symmetries given by the class-$A$ symmetry group $\mathrm{U}(1,1|2)$.

When the physical system is in the metal phase, both $\lambda_\sigma$ and $\lambda_\tau$ flow to infinity under renormalization, thus approaching the RG-fixed point of a Gaussian free field with full symmetry breaking. When the system resides in the Anderson insulator phase,
both stiffness parameters renormalize to zero so as to yield full symmetry restoration. Now, as a striking consequence of the large anisotropy $\lambda_\sigma \gg \lambda_\tau$ a third possibility arises: the small longitudinal stiffness $\lambda_\tau$ may renormalize to zero while the large transverse stiffness $\lambda_\sigma$ flows to infinity. In other words, our field theory may have a third RG-fixed point which is totally attractive but differs from the attractive RG-fixed points for the metal and the insulator. Then on basic grounds, the coarse topology of the resulting RG-flow diagram has to be the one shown in Fig.\ \ref{fig:F3}. Physical systems in the basin of attraction of the third RG-fixed point are expected to exhibit fractal energy eigenstates (more precisely, percolating clusters of finite transverse size) and singular continuous energy spectrum.

In the limit of $\lambda_\sigma \to \infty$, the transverse degrees of freedom become Gaussian free fields. These can be integrated out exactly, with the resulting reduced theory being a so-called principal chiral model. (The same model in 2D, augmented by a Wess-Zumino-Witten term, had been proposed earlier as the conformal field theory describing the scaling limit of the integer quantum Hall transition \cite{CFT-IQHT}.) In a subsequent paper, we plan to show that the reduced theory does exhibit flow to $\lambda_\tau = 0$, by the non-perturbative effect of topological excitations in the fermion-fermion sector of the supersymmetric theory.

As mentioned in the Introduction, there is an irreconcilable conflict between the Anderson-localization nonlinear $\sigma$-model and conformal symmetry. In the light of our present results, we expect this conflict to be resolved by correcting the field-theory analysis, not by declaring violations \cite{KCGM21, KGM22} of conformal symmetry.

\bibliography{Paper-I.bib}
\end{document}